\documentclass[aps,prd,onecolumn,superscriptaddress,nofootinbib,floatfix]{revtex4-2}

\usepackage{amsmath}
\usepackage{amssymb}
\usepackage{slashed}
\usepackage{graphicx}
\usepackage[colorlinks=true,linkcolor=blue,citecolor=blue,urlcolor=blue]{hyperref}

\newcommand{\rtopi}{\tau \to \pi \nu_\tau}
\newcommand{\rtoK}{\tau \to K \nu_\tau}
\newcommand{\rtoKg}{\tau \to K \nu_\tau \gamma}
\newcommand{\rtoPg}{\tau \to P \nu_\tau \gamma}
\newcommand{\Kmng}{K^+ \to \mu^+ \nu_\mu \gamma}
\newcommand{\drsd}{\delta_{rSD}}
\newcommand{\FV}{F_V}
\newcommand{\FA}{F_A}
\newcommand{\fpi}{f_\pi}
\newcommand{\FVDF}{F_V^{\rm DF}}
\newcommand{\FADF}{F_A^{\rm DF}}
\newcommand{\FVGR}{F_V^{\rm GR}}
\newcommand{\FAGR}{F_A^{\rm GR}}

\begin{document}

\title{Comment on ``Radiative corrections to $\tau \to \pi(K)\nu_\tau[\gamma]$:
  A reliable new physics test''}

\author{Markus Finkemeier}
\email{mf.718.622@gmail.com}
\affiliation{Independent researcher, Brooklyn, NY, USA}

\begin{abstract}
Arroyo-Ure\~na, Hern\'andez-Tom\'e, L\'opez-Castro, Roig, and
Rosell~\cite{AHLRR-PRD,AHLRR-JHEP} (AHLRR) take the structure-dependent (SD) amplitude for
$\rtoPg$ from the resonance chiral theory result of Guo and Roig~\cite{GR10} (GR10) and
obtain the real-photon SD correction $\drsd^{\tau\pi} = +0.15\%$. We show that the
dictionary relating the GR10 and Decker-Finkemeier (DF)~\cite{DF93,DF94} form factor
conventions, stated in footnote~2 of Ref.~\cite{GR10} and repeated in footnote~1 of
Ref.~\cite{AHLRR-JHEP}, contains a sign error in its axial entry: the printed
amplitudes of
the two papers imply $\FADF = -2\sqrt{2}\, m_P\, \FAGR$, not $+2\sqrt{2}\, m_P\, \FAGR$.
With the corrected dictionary, the GR10 amplitude agrees with the
$\mathcal{O}(p^4)$ chiral prediction~\cite{BEG93} in the axial sector but carries the
opposite sign to the chiral anomaly~\cite{WZ71,Witten83} in the vector sector, and the
relative sign adopted in the DF addendum~\cite{DF94} is confirmed in both sectors. The
identification of the AHLRR value with the pre-addendum DF result (footnote~6 of
Ref.~\cite{AHLRR-JHEP}) rests on the erroneous dictionary and does not hold. Because the
vector interference integrates to a small value, the numerical impact of the sign
correction is minor: $\drsd^{\tau\pi}$ shifts from $+0.150\%$ to $+0.146\%$ and
$\drsd^{\tau K}$ from $+0.18\%$ to approximately $+0.16\%$. The dominant
uncertainty in $\drsd$ is
instead the form factor shape dependence of the axial interference, which exceeds the
quoted errors.
\end{abstract}

\maketitle

\section{Two conventions and their dictionary}
\label{sec:dictionary}

Both DF~\cite{DF93} and GR10~\cite{GR10} write the amplitude for
$\tau^-(p_\tau) \to \nu_\tau(q)\, P^-(p)\, \gamma(k)$ as the sum of an internal
bremsstrahlung (IB) part, fixed by QED and $F_P$, and an SD part parametrized by a vector
and an axial form factor. The DF form factors are dimensionless; the GR10 form factors
carry dimension of inverse mass. Footnote~2 of Ref.~\cite{GR10} states the translation
\begin{equation}
  \FVDF(t) = \sqrt{2}\, m_P\, \FVGR(t)\,, \qquad
  \FADF(t) = +2\sqrt{2}\, m_P\, \FAGR(t) \qquad \text{[footnote~2 of Ref.~\cite{GR10}]},
  \label{eq:fn2}
\end{equation}
and footnote~1 of Ref.~\cite{AHLRR-JHEP} repeats it, adding that the two conventions
``differ by a global $-i$ factor'' to which the discrepancy between the GR10-based and the
DF-based values of $\drsd$ is attributed. We now show that the axial entry of
Eq.~(\ref{eq:fn2}) has the wrong sign. The demonstration uses only four printed equations,
contains no Levi-Civita tensor, and is independent of all phase, spinor, and projector
conventions.

Take the simplified amplitudes printed in Eq.~(25) of Ref.~\cite{DF93} and Eq.~(25) of
Ref.~\cite{GR10}. In the DF notation ($s \equiv p_\tau$, $\gamma_\pm \propto 1 \pm
\gamma^5$, common weak factors suppressed),
\begin{align}
  \mathcal{M}_{\rm IB}^{\rm DF} &= \fpi\, m_\tau\,
     \bar u(q)\, \gamma_+ \mathcal{B}\, u(s)\,, &
  \mathcal{B} &\equiv \frac{p\cdot\epsilon}{p\cdot k}
     - \frac{s\cdot\epsilon}{s\cdot k}
     + \frac{\slashed{k}\slashed{\epsilon}}{2\, s\cdot k}\,,
  \label{eq:DFIB}\\
  \mathcal{M}_{A}^{\rm DF} &= \frac{1}{\sqrt 2}\, \frac{\FADF(t)}{m_P}\,
     \bar u(q)\, \gamma_+ \mathcal{A}\, u(s)\,, &
  \mathcal{A} &\equiv (p\cdot k)\, \slashed{\epsilon} - (\epsilon\cdot p)\, \slashed{k}\,,
  \label{eq:DFA}
\end{align}
while GR10 write [their Eq.~(25), same external momenta, $i$ on the left of every
amplitude]
\begin{align}
  i\,\mathcal{M}_{\rm IB}^{\rm GR} &= F_P\, M_\tau\,
     \bar u(q)\, (1+\gamma^5) \left[ \frac{p_\tau\cdot\epsilon}{p_\tau\cdot k}
     - \frac{p\cdot\epsilon}{p\cdot k}
     - \frac{\slashed{k}\slashed{\epsilon}}{2\, p_\tau\cdot k} \right] u(p_\tau)
   = -\, F_P\, M_\tau\, \bar u(q)\, (1+\gamma^5)\, \mathcal{B}\, u(p_\tau)\,,
  \label{eq:GRIB}\\
  i\,\mathcal{M}_{A}^{\rm GR} &= \FAGR(t)\,
     \bar u(q)\, (1+\gamma^5) \left[ (t - m_P^2)\, \slashed{\epsilon}
     - 2 (\epsilon\cdot p)\, \slashed{k} \right] u(p_\tau)
   = 2\, \FAGR(t)\, \bar u(q)\, (1+\gamma^5)\, \mathcal{A}\, u(p_\tau)\,,
  \label{eq:GRA}
\end{align}
where the last step in Eq.~(\ref{eq:GRIB}) is sign reordering of the same bracket, and the
last step in Eq.~(\ref{eq:GRA}) uses $t - m_P^2 = 2\, p\cdot k$ for a real photon. Both
papers describe the same physical process with the same IB normalization ($F_P = \fpi$),
so the ratio $\mathcal{M}_A / \mathcal{M}_{\rm IB}$, in which every common factor
(including the $i$ of GR10's left-hand sides and the normalization of $\gamma_\pm$)
cancels, must coincide:
\begin{equation}
  \frac{\FADF}{\sqrt{2}\, m_P\, \fpi\, m_\tau}
  = \frac{2\, \FAGR}{-\, \fpi\, m_\tau}
  \qquad\Longrightarrow\qquad
  \boxed{\;\FADF(t) = -\,2\sqrt{2}\, m_P\, \FAGR(t)\;}\,,
  \label{eq:axialmap}
\end{equation}
opposite in sign to Eq.~(\ref{eq:fn2}).

The vector entry of the dictionary involves the $\varepsilon^{\mu\nu\rho\sigma}$
conventions of the two papers. These are fixed internally: each paper's printed amplitude
reproduces its own printed density functions [Eqs.~(28)--(29) of Ref.~\cite{DF93},
Eqs.~(41)--(42) of Ref.~\cite{GR10}] in exactly one $\varepsilon$ convention, and the two
conventions are opposite. This is in fact recorded by GR10 themselves: their footnote~3
notes that their printed $f_{VA}$ and $f_{IB-V}$ carry the opposite signs to those of DF,
which, given the relations above, is the statement that the two papers use opposite
$\varepsilon^{0123}$. Repeating the ratio argument for the vector amplitudes with this
input confirms the vector entry of Eq.~(\ref{eq:fn2}),
$\FVDF = +\sqrt{2}\, m_P\, \FVGR$. We have verified all of these statements by a numerical
implementation of the printed amplitudes and densities of both papers: each set is
internally consistent, and Eq.~(\ref{eq:axialmap}) holds with zero spread over phase
space, photon polarizations, and fermion spins.

Two consequences follow immediately. First, no single global factor, $-i$ or otherwise,
relates the two conventions: the vector and axial dictionary entries carry opposite signs.
The premise of footnote~1 of Ref.~\cite{AHLRR-JHEP}, that the GR10-DF discrepancy in
$\drsd$ is a convention artifact, fails. Second, the GR10 and DF93 amplitudes are not
physically equal. Throughout, the superscript on $\FVDF$, $\FADF$ and $\FVGR$, $\FAGR$
labels the \emph{convention} in which a form factor is written (the dimensionless DF
normalization of footnote~2 of Ref.~\cite{GR10} and footnote~1 of Ref.~\cite{AHLRR-JHEP},
or the GeV$^{-1}$ GR10 normalization), not the paper a value is taken from. The chiral
limits employed by GR10 and AHLRR, $\FVGR(0) = -N_C/(24\pi^2 \fpi) < 0$ and
$\FAGR(0) < 0$, re-expressed in the DF convention through the corrected dictionary, read
\begin{equation}
  \FVDF(0) = -0.027 < 0\,, \qquad \FADF(0) = +0.012 > 0\,,
  \label{eq:GRinDF}
\end{equation}
the wrong vector sign and the correct axial one. DF93 instead used
$\FVDF(0) = -0.027$ \emph{and} $\FADF(0) = -0.012$ (wrong in both sectors), following the
earlier radiative pion decay literature~\cite{BP68}; the DF addendum~\cite{DF94} reversed
both, recovering the values fixed by the anomaly and by $L_9 + L_{10}$
[Eq.~(\ref{eq:chiral})]. The three configurations are collected in
Table~\ref{tab:configs}. The GR10 amplitude thus agrees with DF93 in the vector
sector and with DF94 in the axial sector. The identification in footnote~6 of
Ref.~\cite{AHLRR-JHEP} of the AHLRR value $+0.15\%$ with the DF93 (``original relative
sign'') value $+0.17\%$ presupposes the equality of the two amplitudes and does not hold;
the numerical proximity of the two values is partly coincidental (Sec.~\ref{sec:numbers}).

\begin{table*}[t]
\caption{\label{tab:configs}Chiral-limit form factors in the two conventions, related by
Eqs.~(\ref{eq:fn2}) and~(\ref{eq:axialmap}). We argue here that the signs in the DF94
addendum are the physical ones.}
\begin{ruledtabular}
\begin{tabular}{l|cc|cc}
Paper
 & \multicolumn{2}{c|}{DF convention (dimensionless)}
 & \multicolumn{2}{c}{GR convention (GeV$^{-1}$)} \\
\cline{2-3}\cline{4-5}
 & $\FVDF(0)$ & $\FADF(0)$ & $\FVGR(0)$ & $\FAGR(0)$ \\
\colrule
DF94 addendum ($=$ WZW $+ L_9 + L_{10}$, Eq.~(\ref{eq:chiral}))
                    & $+0.027$ & $+0.012$ & $+0.137$ & $-0.030$ \\
GR10 / AHLRR        & $-0.027$ & $+0.012$ & $-0.137$ & $-0.030$ \\
DF93 (pre-addendum) & $-0.027$ & $-0.012$ & $-0.137$ & $+0.030$ \\
\end{tabular}
\end{ruledtabular}
\end{table*}

\section{The physical signs}
\label{sec:signs}

At $\mathcal{O}(p^4)$ in chiral perturbation theory the two form factors are fixed with no
free parameters: the vector one by the Wess-Zumino-Witten anomaly
term~\cite{WZ71,Witten83}, the axial one by the low-energy constants $L_9 + L_{10}$. In
the DF (dimensionless, Particle-Data-Group-type~\cite{PDG2024}) convention,
\begin{align}
  \FVDF(0) &= +\frac{N_C\, m_P}{12\sqrt{2}\,\pi^2 \fpi} = +0.027\,, \qquad
  \FADF(0) = +\frac{4\sqrt{2}\, m_P}{\fpi}\, (L_9 + L_{10}) = +0.012\,,
  \nonumber\\
  \gamma &\equiv \frac{\FADF(0)}{\FVDF(0)} = \frac{96\pi^2 (L_9+L_{10})}{N_C}
  \simeq +0.44\,,
  \label{eq:chiral}
\end{align}
for $P = \pi$, using $L_9 = 6.9\times 10^{-3}$, $L_{10} = -5.5\times 10^{-3}$. We have
re-derived both signs directly from the gauged anomaly functional and the
Gasser-Leutwyler $L_9$, $L_{10}$ terms; they agree with the
$K_{\ell 2\gamma}$ results of Bijnens, Ecker, and Gasser~\cite{BEG93} [their Eq.~(3.36),
transported to the present convention through their Eq.~(B.11)]. Both form factors are
positive, with the ratio $\gamma \simeq +0.44$ in agreement with the measured
$\gamma = 0.443 \pm 0.015$ from radiative pion decay~\cite{PIBETA04}; since the measured
$\gamma$ is positive, the experimental determination of either sign fixes the other. An
independent, nonperturbative determination supports this relative sign: using light-cone
sum rules, Bansal and Mahajan~\cite{BM20} obtain $\gamma \simeq +0.47$ for $\rtoPg$ with
$F_V(0)$ fixed by the anomaly, reproducing the DF addendum configuration
[Eq.~(\ref{eq:chiral})] by a method independent of the $\mathcal{O}(p^4)$ chiral counting
used here.

The vector sign has been measured. With both form factors positive, the IB-SD
interference in $\Kmng$ is predicted to be net destructive, dominated by the
$\mathrm{INT}^- \propto (\FV - \FA)$ term: Table~4 of Ref.~\cite{BEG93} gives
$\mathrm{INT}^- = -3.83\times 10^{-5}$ and $\mathrm{INT}^+ = +1.44\times 10^{-5}$ over the
full phase space. ISTRA+~\cite{ISTRA} observed destructive interference in the
structure-dependent region of $\Kmng$ and obtained
$\FV - \FA = 0.21 \pm 0.04 \pm 0.04 > 0$, in agreement with the prediction. Reversing the
relative sign of the SD amplitude would reverse the observed interference.

Comparing with Eq.~(\ref{eq:GRinDF}): the DF94 amplitude carries the correct sign in both
sectors; the GR10 amplitude, and with it the AHLRR analysis, carries the correct axial
sign and the opposite of the correct vector sign; the DF93 amplitude carried the opposite
sign in both sectors. The sign reversal performed in the DF addendum~\cite{DF94}, there
motivated by the chiral results of Ref.~\cite{BEG93} and by SU(3) relations to the
$K_{\ell 4}$ anomalous form factor, is thereby confirmed. Reference~\cite{GR10}
(Sec.~I) attributes that reversal to the absence of a Lagrangian approach in
Refs.~\cite{DF93,DF94}; the present analysis shows the reversal was correct, and that it
is the vector sector of the Lagrangian computation of Ref.~\cite{GR10} that disagrees with
the chiral anomaly.

\section{Numerical consequences}
\label{sec:numbers}

The correction to the AHLRR results is the reversal of $\mathrm{Re}\,\FVGR$, i.e.\ of the
IB-V interference. The V-A interference vanishes identically when integrated over the full
phase space [$d\Gamma_{VA}/dz = 0$, Eq.~(44) of Ref.~\cite{GR10}], and $|\FV|^2$,
$|\FA|^2$ terms are unaffected. From Table~I of
Ref.~\cite{GR10} (pion channel, $E_\gamma \geq 50$~MeV, in units of
$10^{-3}\,\Gamma_{\rtopi}$): $\Gamma_{IB-V} = +0.02$, $\Gamma_{IB-A} = +0.34$,
$\Gamma_{VV} = 0.99$, $\Gamma_{AA} = 0.15$, which sum to $1.50 \times 10^{-3}$, the AHLRR
value $\drsd^{\tau\pi} = +0.15\%$. Reversing the IB-V term gives
\begin{equation}
  \drsd^{\tau\pi} = +0.150\% \;\longrightarrow\; +0.146\%\,, \qquad
  \Delta\drsd^{\tau\pi} = -2\,\Gamma_{IB-V}/\Gamma_{\rtopi} = -0.004\%\,.
\end{equation}
For the kaon channel, the AHLRR value $\drsd^{\tau K} = (0.18 \pm 0.05)\%$ corresponds to
the $c_4 = 0$ column of Table~II of Ref.~\cite{GR10} (the $c_4 = -0.07$ column yields
$\Gamma_{VV} \approx 59 \times 10^{-3}\,\Gamma_{\rtoK}$, flagged as anomalous in
Ref.~\cite{GR10} itself), where
$\Gamma_{IB-V} = +0.10 \times 10^{-3}\,\Gamma_{\rtoK}$, so, approximately,
\begin{equation}
  \drsd^{\tau K} = +0.18\% \;\longrightarrow\; +0.16\%\,, \qquad
  \Delta\drsd^{\tau K} \approx -0.02\%\,,
\end{equation}
within the quoted uncertainty. Propagated to the lepton universality test of
Ref.~\cite{AHLRR-JHEP}, the shifts in $|g_\tau/g_\mu|$ are below $10^{-4}$ in the pion
channel and of order $10^{-4}$ in the kaon channel, negligible against current
experimental errors~\cite{Pich14,HFLAV24}; the impact on $|V_{us}|$ determinations from
$\rtoKg$ is correspondingly small.

The spread between the DF value $\drsd^{\tau\pi} = +0.05\%$~\cite{DF95} and the
corrected value $+0.146\%$ is therefore not a sign effect. It is dominated by the form
factor shape dependence of the axial interference integral, which is sensitive to the
$a_1$ region, where $\mathrm{Re}\,\FA(t)$ changes sign: the resonance chiral theory form
factors give $\Gamma_{IB-A} = +0.34\times 10^{-3}\,\Gamma_{\rtopi}$~\cite{GR10}, the
vector-dominance parametrization of Refs.~\cite{DF93,DF95} gives
$\approx -0.6 \times 10^{-3}\,\Gamma_{\rtopi}$, and constant $\mathcal{O}(p^4)$ form
factors give $-0.24\times 10^{-3}\,\Gamma_{\rtopi}$ (this work), all with the same,
correct, relative signs of the amplitudes. This parametrization dependence, which can
reverse the sign of the integrated interference, exceeds the uncertainty quoted in
Ref.~\cite{AHLRR-JHEP} for $\drsd$ and should be reflected in applications, in particular
in radiative-correction inputs to $\tau$-based lepton universality and $|V_{us}|$
analyses~\cite{HFLAV24}.

\section{Summary}
\label{sec:summary}

(i) The convention dictionary in footnote~2 of Ref.~\cite{GR10} and footnote~1 of
Ref.~\cite{AHLRR-JHEP} has a sign error in its axial entry; the correct relation is
Eq.~(\ref{eq:axialmap}). No global factor relates the GR10 and DF conventions.
(ii) Consequently the GR10 amplitude does not carry the pre-addendum DF93 relative sign:
it agrees with DF94 in the axial sector, and footnote~6 of Ref.~\cite{AHLRR-JHEP} does not
hold.
(iii) The relative sign adopted in the DF addendum~\cite{DF94} is the physical one in both
sectors, fixed by the chiral anomaly and $L_9 + L_{10}$ and anchored by the ISTRA+
measurement~\cite{ISTRA}; the vector sector of the GR10 amplitude carries the opposite
sign and should be corrected.
(iv) The numerical impact on $\drsd$ is small, $-0.004\%$ ($\pi$) and $\approx -0.02\%$
($K$); the dominant uncertainty in $\drsd$ is instead the form factor shape dependence of
the axial interference, which is larger than the quoted errors.

\begin{acknowledgments}
The author used Claude (Anthropic) for assistance with the technical analysis, literature
search, and drafting. The structure, framing, and conclusions are the author's. The author
takes responsibility for the contents.
\end{acknowledgments}

\bibliography{comment_v19}

\end{document}